\documentclass[aps, prd, superscriptaddress, twocolumn,a4paper]{revtex4-1}

\usepackage{hyperref}
\usepackage{graphicx}
\usepackage{amsmath}
\usepackage{color}
\usepackage{float}
\usepackage{bm}
\usepackage[english]{babel}

\newcommand{\be}{\begin{equation}}
	\newcommand{\ee}{\end{equation}}
\newcommand{\ba}{\begin{eqnarray}}
	\newcommand{\ea}{\end{eqnarray}}

\begin{document}
	
	\title{ $c {\bar c}$ and $b {\bar b}$ suppression in Glasma}
	
	\author{Pooja}\email{pooja19221102@iitgoa.ac.in}
	\affiliation{School of Physical Sciences, Indian Institute of Technology Goa, Ponda-403401, Goa, India}
	
	\author{Mohammad Yousuf Jamal}\email{yousufjml5@gmail.com}
	\affiliation{School of Physical Sciences, Indian Institute of Technology Goa, Ponda-403401, Goa, India}
	
	\author{Partha Pratim Bhaduri}\email{partha.bhaduri@vecc.gov.in}
	\affiliation{Variable Energy Cyclotron Centre, Homi Bhabha National Institute, Kolkata, India}
	
	\author{Marco Ruggieri}\email{marco.ruggieri@dfa.unict.it}
	\affiliation{Department of Physics and Astronomy "Ettore Majorana", University of Catania, Via S. Sofia 64, I-95123 Catania, Italy}\affiliation{INFN-Sezione di Catania, Via S. Sofia 64, I-95123 Catania, Italy}

	\author{Santosh K. Das}\email{santosh@iitgoa.ac.in}
	\affiliation{School of Physical Sciences, Indian Institute of Technology Goa, Ponda-403401, Goa, India}

	% ************************* Start : Abstract ******************************* %
\begin{abstract}
This study investigates the evolution and dissociation dynamics of $c\bar{c}$ and $b\bar{b}$ pairs within the 
pre-equilibrium, gluon-dominated stage of high energy
nuclear collisions.
An attractive potential made of a perturbative Coulomb-like term and of a confining term is used to 
simulate the attractive strong force in the 
pairs.
Besides, we implement the interaction of the pairs
with the evolving Glasma fields by virtue of the
Wong equations.
The interaction with the classical color fields
dominates the dynamics,
causing an increase in pair separation and subsequent dissociation. The observed finite probability of dissociation for these states reveals the intricate interplay between QCD dynamics and the suppression of $c\bar{c}$ and $b\bar{b}$ states during the pre-equilibrium stage. The research highlights differences between $c\bar{c}$ and $b\bar{b}$ pairs, revealing the role of quark flavor in the dissociation process. Dissociation spectra analysis indicates a peak shift towards higher momentum, reflecting a slight energy gain by the pairs. This investigation provides valuable insights into the complex dynamics of $c\bar{c}$ and $b\bar{b}$ pairs in the Glasma, which may help in better interpretation of experimental results on further integration with subsequent phases of the created matter.

		%In high-energy heavy-ion collision experiments, the study of the pre-thermalization phase known as the Glasma and its subsequent evolution into the Quark-Gluon Plasma has revealed profound insights into Quantum Chromodynamics. The Glasma, arising from the collision of ultra-relativistic heavy ions, is characterized by highly non-equilibrium color fields and elevated energy densities, significantly influencing the early-stage dynamics of the collision system. Within this context, heavy quarks and their  states being unique probes play a crucial role in providing valuable insights into the properties of the evolving QCD medium and early-time dynamics. We, for the first time, investigated the interaction of  $q\bar{q}$ states with the Glasma phase, particularly $c\bar{c}$ and $b\bar{b}$, and observed that there is a finite probability of dissociation of these states.  This study unveils the complex interplay between QCD dynamics and the suppression of  $q\bar{q}$ states, contributing to our comprehensive understanding of heavy-ion collisions.
	\end{abstract}
	% ************************* End : Abstract ******************************* %

	% ~~~~~~~~~~~~~~~~~~~~~~~~~~~~~~~~~~~~~~~~~~~~~~~~~~~~~~~~~~~~~~~~~~~~~~~~ %
	\maketitle
	% ~~~~~~~~~~~~~~~~~~~~~~~~~~~~~~~~~~~~~~~~~~~~~~~~~~~~~~~~~~~~~~~~~~~~~~~~ %

	% ######################## Start : Introduction ########################  %
	\section{Introduction}
	\label{sec:intro}
	
	Relativistic heavy-ion collision (HIC) experiments, conducted notably at the Relativistic Heavy Ion Collider (RHIC) and the Large Hadron Collider (LHC), serve as a crucial platform for investigating the pre-equilibrium phase termed the Glasma~\cite{Kovner:1995ja, Kovner:1995ts, Gyulassy:1997vt, Lappi:2006fp, Krasnitz:2003jw, Fukushima:2006ax, Fujii:2008km, Fukushima:2013dma, Romatschke:2005pm, Romatschke:2006nk, Fukushima:2011nq}. This exploration extends to the subsequent evolution towards the equilibrated phase, commonly referred to as the Quark-Gluon Plasma (QGP)~\cite{Shuryak:2004cy, Jacak:2012dx}, making this field captivating within heavy-ion phenomenology. The Glasma emerges in the early stage following the relativistic collision of two heavy ions carrying a significant number of protons and neutrons, such as $Au-Au$ or $Pb-Pb$ nuclei in high-energy HICs. Before the collision, high-energy protons and neutrons within the nuclei transform into a dense system of gluons and quarks~\cite{McLerran:1993ni, McLerran:1993ka, McLerran:1994vd, Gelis:2010nm, Iancu:2003xm, McLerran:2008es, Gelis:2012ri}. This leads to the creation of two sheets of colored glass, generating strong longitudinal gluonic fields within the forward light cone, known as the Color Glass Condensate (CGC). The CGC represents the saturation of gluon fields in the nuclei, resulting in a high density of color charges.
	Upon the overlap of the two nuclei, intense gluon fields give rise to the formation of color flux tubes connecting colliding nucleons. These tubes, characterized by highly concentrated color fields, constitute the pre-equilibrium phase. This phase is marked by its classical nature, with a high intensity denoted as $A_\mu^a \simeq Q_s/g$, where $Q_s$ is the saturation scale, and $g$ is the coupling constant of Quantum Chromodynamics (QCD). The time evolution of these gluon fields is governed by Classical Yang-Mills (CYM) equations. 
	The collision of color flux tubes culminates in the eventual formation of the Glasma. Defined as a phase characterized by strong, highly non-equilibrium color fields and elevated energy densities, the Glasma triggers rapid initial entropy production. These factors exert a significant influence on the non-equilibrium dynamics and evolution of strong color fields, playing a central role in shaping the overall system dynamics. Despite its non-trivial and highly energetic nature, current research in high-energy nuclear physics is dedicated to comprehending the formation and evolution of the Glasma.
	In this context, the evolving fields during the Glasma phase are denoted as EvGlasma. The duration of the EvGlasma phase corresponds to the pre-hydro stage and typically spans around $1~\mathrm{fm/c}$ for $AA$ collisions at RHIC energy and approximately $0.3~\mathrm{fm/c}$ for $AA$ collisions at LHC.

	The heavy quarks, generated during the early stages of high-energy HICs, emerge as essential probes to study the QGP phase~\cite{Prino:2016cni,Andronic:2015wma,Rapp:2018qla, Cao:2018ews, Aarts:2016hap, Greco:2017rro, Dong:2019unq, Das:2024vac, Xu:2018gux, Moore:2004tg, vanHees:2005wb, vanHees:2007me, Gossiaux:2008jv, He:2011qa, Song:2015sfa, Alberico:2011zy, Lang:2012cx, Xu:2017obm, Cao:2016gvr, Das:2016cwd, Das:2015ana, Das:2017dsh, Das:2015aga, Song:2019cqz, Beraudo:2015wsd, Das:2013kea, Scardina:2017ipo, Cao:2015hia, Das:2016llg, Ruggieri:2022kxv, Pooja:2023gqt}. Their production occurs shortly after the collision, with a production time on the order of $\tau_{form} \approx 1/(2M)$, where $M$ represents the mass of heavy quarks. As a non-equilibrium probe and produced early, heavy quarks — like charm and beauty quarks — can provide valuable insights into both the pre-equilibrium and QGP phases. They traverse nearly all stages of HICs, making them excellent indicators of the dynamics involved. Previous calculations typically neglected the momentum evolution of heavy quarks during the pre-equilibrium phase, assuming its duration to be very short.
   In recent years, several efforts have been made to investigate the pre-equilibrium phase using heavy quarks \cite{Mrowczynski:2017kso, Ruggieri:2018rzi, Sun:2019fud, Liu:2019lac, Liu:2020cpj, Boguslavski:2020tqz, Khowal:2021zoo, Pooja:2022ojj, Boguslavski:2023fdm, Carrington:2020sww, Carrington:2022bnv, Avramescu:2023qvv, Pandey:2023dzz} and jets \cite{Ipp:2020nfu, Carrington:2021dvw, Boguslavski:2023alu} as probes. Recent studies suggest that heavy quarks undergo significant diffusion in the Glasma phase \cite{Ruggieri:2018rzi, Sun:2019fud, Liu:2019lac, Liu:2020cpj, Khowal:2021zoo}, with its implications for experimental observations shown in ref.~\cite{Sun:2019fud}. In this context, it would be intriguing to examine the evolution and dissociation dynamics of heavy quarkonia, which are bound states of heavy quarks and anti-quarks ($q\bar{q}$), during the Glasma phase. The survival or dissociation of the pairs depends on their interaction with the medium they traverse. Hence, examining their interactions with the Glasma phase can provide valuable insights into the early-time dynamics of HICs, thereby aiding in a deeper understanding of experimental observations.

	In the context of QGP, the dissociation of heavy quarkonia ($q\bar{q}$), offers deeper insights and has been extensively studied~\cite{Matsui:1986dk, McLerran:1986zb,Srivastava:2012pd, Mocsy:2004bv, Agotiya:2008ie, Datta:2003ww, Brambilla:2008cx, Burnier:2009yu, Dumitru:2009fy, Laine:2006ns, Singh:2015eta, Andronic:2015wma, Rapp:2017chc, Kumar:2023wvz,He:2021zej,Villar:2022sbv,Song:2023ywt,Das:2022lqh,Brambilla:2024tqg,Andronic:2024oxz, Casalderrey-Solana:2012yfo, Brambilla:2016wgg, Akamatsu:2020ypb, Yao:2021lus, Miura:2022arv}. Though the dissociation was originally attributed to Debye color screening of the confining potential in a partonic medium, subsequent theoretical studies revealed a considerable suppression due to the medium-induced imaginary potential of the quarkonia system. In relativistic transport models describing heavy-ion collisions, the effect is realized through collisional dissociation by hard gluons, leading to thermal broadening of the width of the in-medium quarkonium spectral function (see Ref.~\cite{Rothkopf:2019ipj} for the detailed review of the recent studies and progress on this topic. In literature, quarkonia production in hadronic collisions is generally modeled as a factorizable two-state process. The first stage is the perturbative production of the initially compact $q{\bar q}$ pair. This is followed by the subsequent non-perturbative formation of $q\bar{q}$ bound states with appropriate quantum numbers. To date, no first principle calculation can account for the resonance binding and can be treated only phenomenologically~\cite{bodwin1,bodwin2,bodwin3}.
	This dual nature of quarkonia permits us to investigate both perturbative and non-perturbative QCD effects. However, the existence of the quarkonia-bound states in the Glasma phase is a point of investigation. Nevertheless, the Glasma phase must have a significant impact on  $q{\bar q}$ pair that will eventually affect the formation of quarkonia as well as other (D/B)-meson states.  Here, we consider that the states of $q{\bar q}$ exist in the Glasma and may get affected while passing through this phase. It may play a pivotal role in determining the initial conditions that govern quarkonia and (D/B)-meson production, ultimately influencing their yields and properties in the later stages of the collision. 
	
	The remaining part of the manuscript is organized as follows. In Section~\ref{sec:formalism}, we provide the formalism for the evolution of background Glasma and the propagation of the $q {\bar q}$ pairs in it. Section~\ref{sec:results}, is dedicated to the results and discussion. In Section~\ref{sec:Summary_Conclusions}, we summarize the present work and discuss future possibilities. 	
	% ######################## End : Introduction ########################  %

	% ++++++++++++++++++++++++++ Start : Formalism ++++++++++++++++++++++++++ %
\section{Formalism}
\label{sec:formalism}
\subsection{Evolution of background Glasma}
The McLerran-Venugopalan (MV) model, based on the color-glass condensate (CGC) effective field theory, describes colliding objects as Lorentz-contractions of colored glass sheets containing fast partons, which act as sources for slow gluon fields in the saturation regime \cite{Lappi:2007ku}. 

The color charge densities deposited on colliding nuclei $A$ and $B$ are represented as random variables $\rho^a_{A}$ and $\rho^a_{B}$ respectively, following Gaussian statistics as per the following correlation functions 
\begin{eqnarray}
	\langle \rho^a_{A,B}(\textbf{\textit{x}}_T) \rangle&=&  0, \\
	\langle \rho^a_{A,B}(\textbf{\textit{x}}_T)\rho^b_{A,B}(\textbf{\textit{y}}_T)\rangle &=&  (g\mu_{A,B})^2  \delta^{ab} \delta^{(2)}(\textbf{\textit{x}}_T - \textbf{\textit{y}}_T).
	\label{eq:colorchag88}
\end{eqnarray}
where $a$ denotes the adjoint color index. Limiting ourselves to the $SU(2)$ QCD theory, $a$ can take values $1,2,3$ only. The density of color charges in the transverse plane is denoted by $\mu$, hence $g\mu_{A,B}$ is the color charge density and $g\mu = O(Q_s)$ \cite{Lappi:2007ku},
where $Q_s$ is the saturation momentum. $Q_s$ is the only relevant energy scale for the Glasma.

To compute the Glasma fields,  firstly, we solve the Poisson's equations given as
\begin{equation}
	-\partial^2_{\perp}\Lambda^{(i)}(\textbf{\textit{x}}_T)
	=\rho^{(i)}(\textbf{\textit{x}}_T),
\end{equation}
where $i=A, B$. The color charge distribution of the $i^{th}$ nuclei generates the gauge potentials. Further, we compute the Wilson lines as 
\begin{eqnarray}
	V^\dagger(\textbf{\textit{x}}_T) &=& e^{-ig\Lambda^{(A)}(\textbf{\textit{x}}_T)},\\
	W^\dagger(\textbf{\textit{x}}_T) &=& e^{-ig\Lambda^{(B)}(\textbf{\textit{x}}_T)}.
\end{eqnarray}

Hence, the gauge fields of the two colliding objects are evaluated as
\begin{eqnarray}
	\alpha_i^{(A)} &=& \frac{-i}{g}V\partial_iV^\dagger,
	\label{eq:alpha1pp}\\
	\alpha_i^{(B)} &=& \frac{-i}{g}
	W\partial_iW^\dagger.\label{eq:alpha2pp}
\end{eqnarray}

In terms of these gauge fields, at the initial time, the Glasma gauge potential is described as
\begin{equation}
	A_i = \alpha_i^{(A)} +\alpha_i^{(B)},~~~A_z=0,
\end{equation}
where $i = x, y$ and $z$ is considered as the direction of the flight of two nuclei. 
The initial transverse color fields vanish, while the longitudinal Glasma fields at an initial time are given as
\begin{eqnarray}
	E^z &=& -ig \sum_{i=x,y}\bigg[\alpha^{(B)}_i, \alpha^{(A)}_i\bigg],\\
	B^z &=& -ig \bigg(\bigg[\alpha^{(B)}_x, \alpha^{(A)}_y\bigg]+\bigg[\alpha^{(A)}_x, \alpha^{(B)}_y\bigg]\bigg),
\end{eqnarray}	

Once initializing the Glasma gauge potential, Glasma color electric and magnetic fields, we aim to evolve the background Glasma by means of the Classical Yang-Mills equations, namely
\begin{eqnarray}
	\frac{dA^a_i(x)}{dt}&=& E^a_i(x),\\
	\frac{dE^a_i(x)}{dt}&=& \partial_jF^a_{ji}(x)+g f^{abc} A^b_j(x)F^c_{ji}(x),
\end{eqnarray}
where $f^{abc} = \varepsilon^{abc}$ with $ \varepsilon^{123}=+1$. 
Using the standard Einstein summation convention, we define the magnetic part of the field strength tensor as
\begin{equation}
	F^a_{ij}(x) = \partial_i A^a_j(x)- \partial_j A^a_i(x)+ g f^{abc} A^b_i(x)A^c_j(x)
\end{equation}

\subsection{Propagation of the $q {\bar q}$ pairs in the Evolving Glasma} 
The heavy quarks, $c$ and $b$, are produced by hard QCD scatterings at the onset of relativistic heavy-ion collisions within a very short time, 
around 0.02-0.06 fm/c. 
Their motion in the EvGlasma is governed by the Wong equations\cite{Ruggieri:2018rzi,Sun:2019fud, Liu:2019lac, Wong:1970fu, Heinz:1984yq}
\begin{eqnarray}
	\frac{dx^i}{dt} &=& \frac{p^i}{E},\\
	\frac{dp^i}{dt} &=& g Q_a F_a^{i\nu}\frac{p_{\nu}}{E} - 
	\frac{\partial V}{\partial x_i}, \label{eq:wong2}\\
	E\frac{dQ_a}{dt} &=& g \varepsilon_{abc}A^\mu_b   
	 p_\mu Q_c,\label{eq:wong3}
\end{eqnarray}
where $i=x,y,z$ and $E=\sqrt{\textbf{p}^2 + M^2}$; 
$Q_a$ represents the 
color charge of the heavy quark, with $a=1, 2, \dots,N_c^2-1$. 
The term on the right-hand side of \eqref{eq:wong2} is the 
relativistic force experienced by the heavy quarks, 
which is made of two contributions.
The first addendum on the right-hand side of the equation
corresponds to a non-abelian version of the electromagnetic Lorentz force.
The second addendum takes into account the 
{attractive} potential
between $q$ and $\bar q$, that we model as
\begin{equation}
	V(r)= -\frac{3\alpha_s  }{4r}+\sigma r,
	\label{eq:prima_1}
\end{equation}	
where $r$ measures the distance between $q$ and $\bar q$.
The first addendum in the right hand side of
Eq.~\eqref{eq:prima_1} denotes 
the strength
of the perturbative short-range contribution, 
while the second one corresponds to the 
long-range non-perturbative
potential. 
The factor of $\frac{3}{4}$ in the Coulombic term corresponds to the color factor of QCD, which is obtained by $C_F = \frac{N_c^2 -1}{2N_c}$. Considering the $SU(2)$ theory in our formulation, $C_F = \frac{3}{4}$.
Moreover, 
$\alpha_s=g^2/4\pi$ where $g$ denotes the QCD
coupling,
and
$\sigma=0.18$ GeV$^2$ is the string tension
which in turn governs the strength of the 
non-perturbative term.

\section{Results and Discussions}
\label{sec:results}
In this section, we present our results on the dissociation of $q{\bar q}$ pair in the 
background of the evolving  Glasma. 
For what concerns the initialization of the $q\bar q$ pairs
we proceed as follows.
We create the 
pairs 
at proper time $\tau_{form} = 1/(2 M)$,
where $M$ is the heavy quark mass. 
We consider a box of transverse-length
$L=4$ fm.
After randomly fixing  the position, 
$r_q=(x_q, y_q, z_q)$,
of a $q$
inside the box,
we 
randomly choose the distance between the $q$
and the  $\bar q$, which we call $r_B$,   
in the range $(0.01-0.5)$ fm.
Then, the position of the $\bar q$,
$r_{\bar q}=(x_{\bar q}, y_{\bar q}, z_{\bar q})$,
is randomly chosen in such a way
\begin{equation}
	r_B = \sqrt{(x_q-x_{\bar q})^2+(y_q-y_{\bar q})^2+(z_q-z_{\bar q})^2 }. 
	\label{eq:theremainingone} 
\end{equation}
Particularly, 
the orientation of ${\bar q}$ with respect to $q$ is obtained by a random extraction of the polar and azimuthal coordinates
of the $\bar q$, then its radial coordinate is
fixed by virtue of Eq.~\eqref{eq:theremainingone}.
We checked that using hydrogen wave 
functions to initialize the positions of the
$q$ and the $\bar q$ does not lead to significant
changes in the results that we discuss in this
study, therefore we limit ourselves to the 
simple coordinates initialization discussed above,
leaving a more refined one to future works.

Next, we turn to momenta initialization.
Generally speaking, the total momentum of the pair is
\begin{equation}
	{\bf p}_q + {\bf p}_{\bar q}={\bf p},\label{eq:p_q}
\end{equation}
where 
${\bf p}_q$ and ${\bf p}_{\bar q}$ denote the
momenta of $q$ and $\bar q$ respectively;
in this work, we limit ourselves to consider
${\bf p}=0$, as well as to the condition $p_{qz}=p_{\bar qz}=0$,
to mimic the mid-rapidity region
of heavy ion collisions.

The transverse momentum distribution of $q$ is initialized within the  
Fixed Order + Next-to-Leading Log (FONLL) QCD result
that reproduces the D-mesons and B-mesons spectra in $pp$ collisions after fragmentation~\cite{FONLL, Cacciari:2012ny}  
\begin{equation}
	\left.\frac{dN}{d^2 p_T}\right|_\mathrm{prompt} = \frac{x_0}{(x_1 + x_3{p_T^{x_1})}^{x_2}};\label{eq:FFNLO}
\end{equation}
the parameters that we use in the calculations are $x_0 =
20.2837$, $x_1 = 1.95061$, $x_2 = 3.13695$ and $x_3 = 0.0751663$ for charm quark and $x_0 = 0.467997$, $x_1 = 1.83805$, $x_2 = 3.07569$ and $x_3 = 0.0301554$ for beauty quark; the slope of the spectrum has been calibrated to a collision at $\sqrt{s}=5.02$ TeV. 
After $\bm p_q$ is known,
the momentum of the companion antiquark
is obtained by~\eqref{eq:p_q} with $\bm p=0$.

Finally, the color charges are initialized on a $3-$ dimensional sphere 
with radius one; $Q_a Q_a$ corresponds to the total squared color charge, which is conserved in the evolution.

The numerical implementation of the potential~\eqref{eq:prima_1}  
is problematic when $r\rightarrow 0$, due to the Coulomb-like term that
diverges for small $r$. In order to numerically 
tackle the divergence of this term, 
we replace~\eqref{eq:prima_1} in the code  
with
\begin{equation}
	V(r)= -\frac{3\alpha_s  }{4r}\left(1-e^{-Ar}\right)+\sigma r,
	\label{eq:prima_1_mod}
\end{equation}
with $A$ being a constant with dimensions of an inverse length.
We verified that changing $A$ within a reasonable range
(particularly,
we take $1/A$ much smaller than the initial size of the $q\bar q$ pair)
does not substantially affect our results. In this calculation, we use A= 1 GeV.

For the choice of the parameters in the glasma sector, we proceed as 
follows.
We first fix the saturation scale, $Q_s$, 
probing several choices in the range $(1,3)$ GeV.
We then use the well-known
one-loop QCD $\beta-$function to  compute $\alpha_s$
at the scale $Q_s$, and from this we compute the QCD coupling $g^2= 4\pi\alpha_s$. 
The value of $g$ is used
to get the color charge densities of colliding nuclei, $\mu \approx Q_s/0.6 g^2$ \cite{Lappi:2007ku}, to be used in Eq.~\eqref{eq:colorchag88}.

\subsection{Dissociation percentage}
\label{sec:percentag}
Using the aforementioned initial setup, we monitor the 
real-time evolution of the pairs within the background Glasma, and observe their dissociation under various scenarios. These dissociated pairs may combine to form D/B-mesons, while the pairs that survived may eventually form a bound state before they enter the QGP phase. Our criterion for determining whether a pair dissociates or survives is based on the final distance between the corresponding evolving pairs in the Glasma. If the separation between the $q {\bar q}$ pair after evolution time, $\tau$ is greater than or equal to a specified cutoff distance, $r_{c}$, we classify it as a dissociated pair; otherwise, it is considered as the survived one. 

{\color{black} Our dissociation criterion can be understood as follows: we assume various possible separation values, $r_c$, at which the force between the quark and antiquark effectively vanishes. In this case, the potential behaves as \cite{Islam:2020bnp},
\ba
V(r) = 
\begin{cases}
   V(r) & \text{if } r < r_c, \\
   V(r_c) & \text{if } r \geq r_c.
\end{cases}
\label{eq:vr}
\ea
Hence,  at $r_c$, the potential becomes constant, indicating that the force between the quark and antiquark drops to zero. An alternative criterion for the dissociation condition could be considered, such as:
\ba
  \mathrm{kinetic~energy} + \mathrm{potential ~energy} > 0.
  \label{eq:peke}
\ea
However, this condition would be more relevant if the potential energy approached zero at large distances. Given that the confining nature of the Cornell potential prevents this, we will proceed with Eq.\eqref{eq:vr} in the current analysis. The criterion in Eq.\eqref{eq:peke} may be explored in future work to further refine the current understanding.
}

\begin{figure}[t!]
		\centering
		\includegraphics[scale=.5]{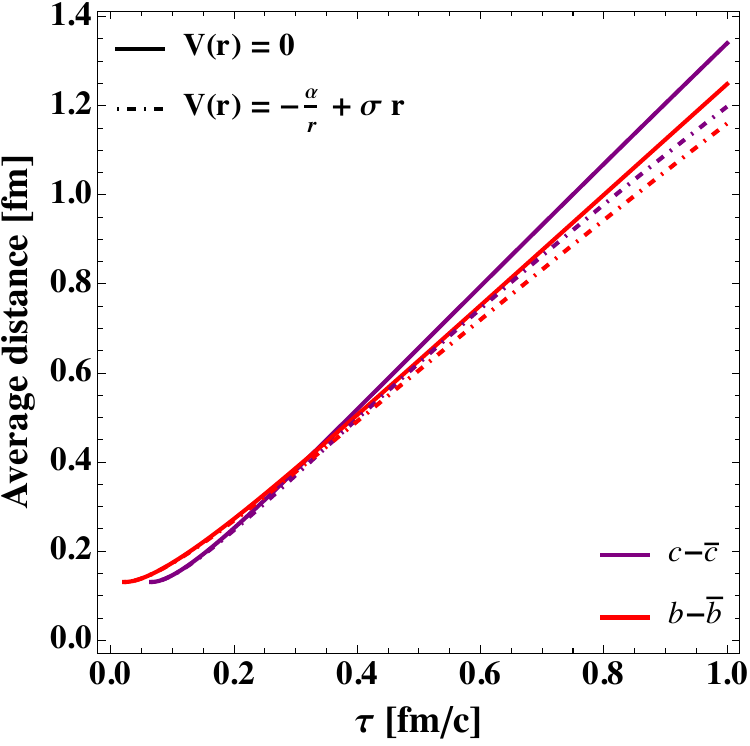}
		\caption{Average distance between 
  quarks and antiquarks in 
  $c{\bar{c}}$ (indigo lines) and $b{\bar{b}}$ (red lines) pairs versus
  time, with (dot-dashed lines) 
  and without (solid lines) the confining potential~\eqref{eq:prima_1}.}
		\label{Fig:separation}
	\end{figure}

As a preamble to what we show later, 
in Fig.\ref{Fig:separation} we plot the
average distance between 
quarks and antiquarks in 
$c{\bar{c}}$ (indigo lines) and $b{\bar{b}}$ (red lines) pairs versus
time, with (dot-dashed lines) 
and without (solid lines) the confining potential~\eqref{eq:prima_1}.
It is interesting to note that $V(r)$ slightly slows down the
coordinates broadening of the $q\bar q$ pair, as it should be since
it provides a {strong attractive force} among $q$ and $\bar q$ in the pair;
however, we note that the effect of the strong gluon fields in which
the quarks diffuse is stronger, and this tends to push the $q$ and the 
$\bar q$ apart.

\begin{figure}[t!]
\includegraphics[scale=.5]{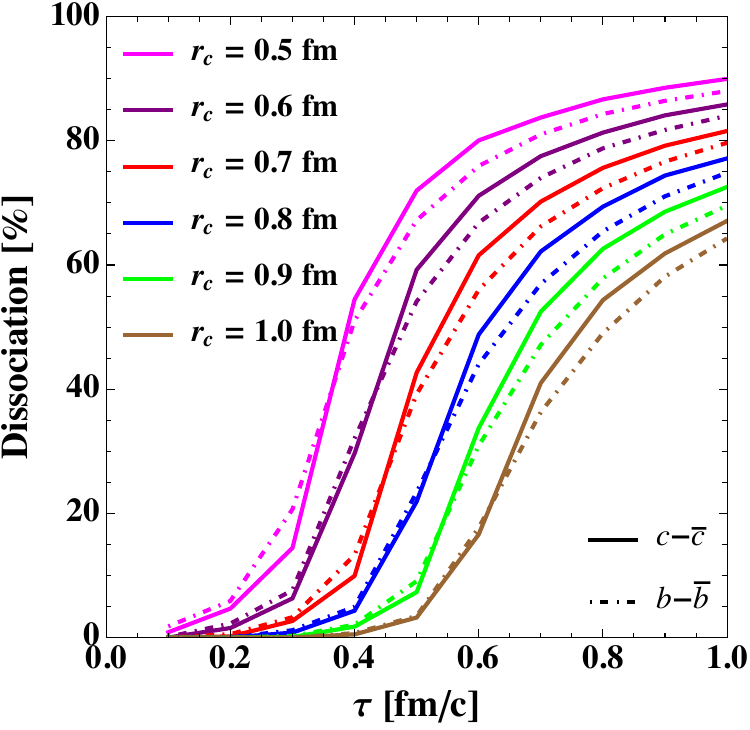}
\caption{Dissociation percentage 
of the $c{\bar{c}}$ (solid curve) and $b{\bar{b}}$ (dot-dashed) pairs 
versus proper time at different cutoff distances.\label{Fig:diss_vs_t}}	
\end{figure}

In Fig.~\ref{Fig:diss_vs_t} we plot 
the percentage of dissociated pairs,
$100\times N_d/N$, where $N_d$ denotes the number of dissociated pairs
while $N$ is the number of total pairs versus time.
In the figure, we show the results for both 
for $c\bar{c}$ (solid lines) and $b\bar{b}$ (dot-dashed lines) 
pairs, 
for specific values of $r_c$. 
In order to estimate $N_d$, we follow the trajectories of $q$ and
$\bar q$ for each pair and  then compute  
the distance between $q$ and $\bar q$ for that pair:
when the distance 
becomes larger than $r_c$, we label that as a dissociated pair. 
We opted for $r_c \ge 0.5$ fm because we generated pairs up to $r_B = 0.5$ fm. 

In Fig.~\ref{Fig:diss_vs_t}
we explore a time range up to $1$ fm/c for illustration purposes,
although the typical lifetime of the 
initial strong gluon fields
is approximately of 0.3-0.6 fm/c. 
We have selected several values of $r_c$,
namely
$r_c =0.5$ fm (magenta lines), $0.6$ fm (violet lines), $0.7$ fm (red lines), $0.8$ fm (blue lines), $0.9$ fm (green lines), $1.0$ fm (brown lines). We observe that the larger the proper time
the
higher the percentage of the dissociated pairs. This is attributed to the growing separation between the $q$ and the $\bar{q}$ in the pairs in the 
early stage, as shown in Fig.\ref{Fig:separation}: 
larger evolution times result in increased distances between $q$
and $\bar{q}$, 
allowing more pairs to have a distance larger than $r_c$ within a finite
amount of time.
The case $r_c = 0.5$ fm exhibits the highest
percentage of dissociated pairs at a specific time, 
whole this percentage decreases when
$r_c$ is increased. 
Qualitatively, we observe a similar pattern for $b\bar{b}$ and $c\bar{c}$
pairs; however, since $b$ quarks have a larger mass, 
they experience a smaller coordinate broadening as compared to charm quarks~\cite{Liu:2020cpj, Khowal:2021zoo};
as a consequence,
we observe less dissociation for $b$ quarks. 
However, the mass difference effect is numerically 
small in all the range of
$r_c$ and $\tau$ studied, so we conclude that
the dissociation probability in the evolving Glasma fields is not very
sensitive to the mass of the heavy quarks.
Remarkably, we find 
a significant range of dissociation probabilities, 
in the range
$60 - 80 \%$  for $\tau = 1.0$ fm/c.

\begin{figure}[t!]
		\includegraphics[scale=.5]{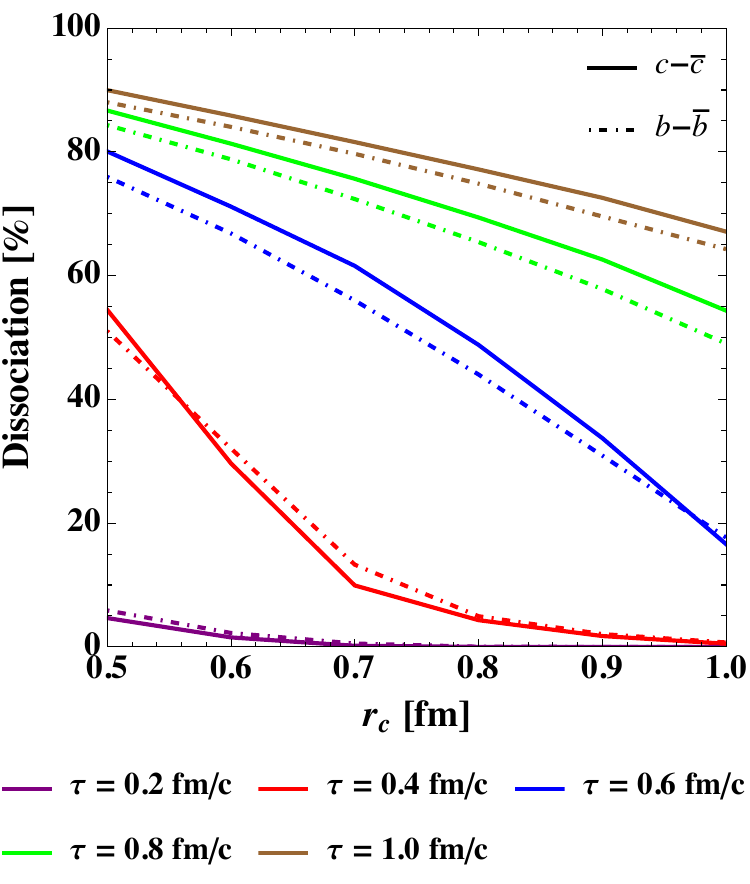}
		\caption{Dissociation percentage
of $c{\bar{c}}$ (solid) and $b{\bar{b}}$ (dashed) versus $r_c$
for several values of the proper time.}
		\label{Fig:diss_vs_r}
	\end{figure}

In order to provide a clearer understanding 
of the impact of $r_c$, we plot the dissociation probability of 
$q\bar{q}$ pairs versus $r_c$, 
at specific values of proper time 
in Fig.\ref{Fig:diss_vs_r}. 
We consider the evolution of the pairs for several values of proper times, namely
$\tau =0.2$ fm/c (violet lines), $0.4$ fm/c (red lines), 
$0.6$ fm/c (blue lines), $0.8$ fm/c (green lines), 
$1.0$ fm/c (brown lines).
The results for $b\bar{b}$ pairs (dot-dashed lines) are qualitatively
similar to those for the $c\bar{c}$ pairs (solid lines).
Quantitatively, the results for $c\bar{c}$ and $b\bar{b}$ pairs are in
the same ballpark.
For any $\tau$, the dissociation probability of the pairs 
decreases as $r_c$ is increased, in agreement with
our criterion for dissociation.
In fact, the coherent gluon fields tend to increase the separation between the $q$ and the $\bar q$ of the pairs: for a particular evolution time, 
there will be a large number of pairs that can beat the lower 
cutoff distance and go farther away from this critical value. 
However, 
we notice that even for the highest $r_c$  considered,
that is $r_c = 1$ fm, for $\tau \sim 0.8$ fm/c we find 
almost the $60 \%$ of the pairs are dissociated
Moreover,
at $r_c > 0.9$ and $\tau\le 0.6$ fm/c, the results for beauty and charm almost overlap, indicating that the effect of mass is not substantial 
in the very early stage.

\subsection{Dissociation Spectra}
\label{sec:spectra}

\begin{figure}[t!] 
		\includegraphics[scale=.5]{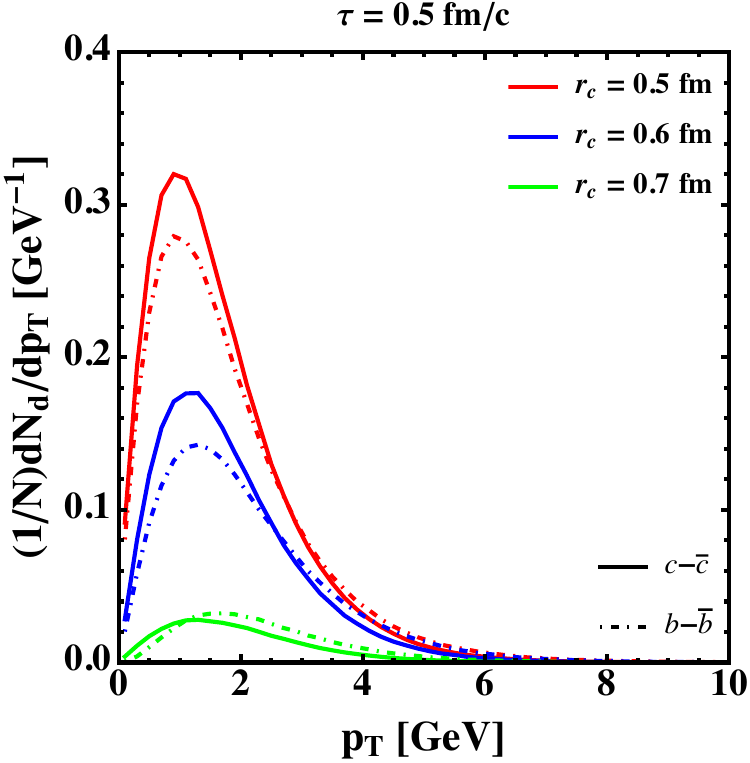}
		\caption{Variation of $(1/N)dN_{d}/dp_T$ for $c{\bar{c}}$ (solid) and $b{\bar{b}}$ (dashed) versus their dissociation momenta at fixed evolution time, $\tau=0.5$ fm/c.}
		\label{Fig:spectra_05}
	\end{figure}

\begin{figure}[t!] 
		\includegraphics[scale=.5]{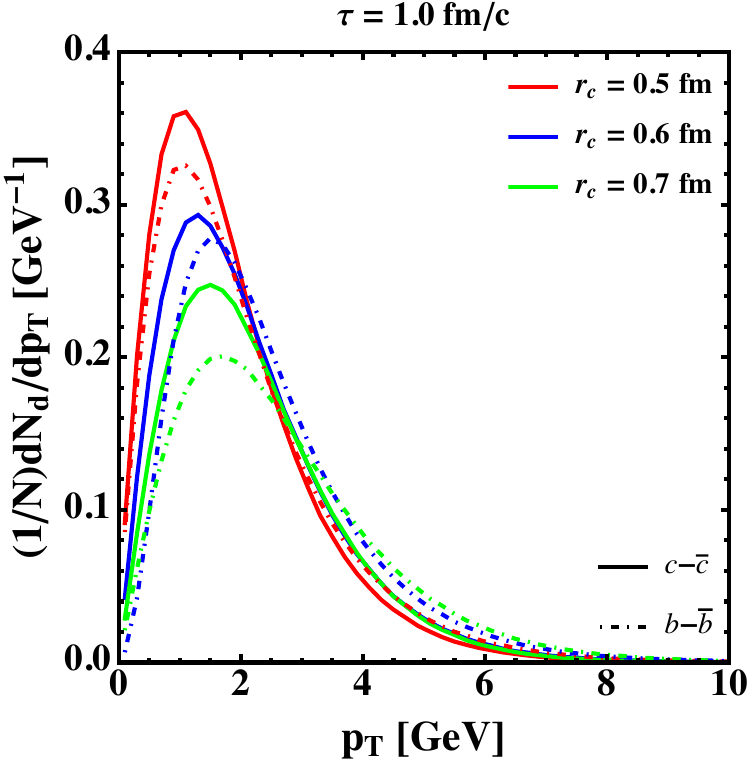}
		\caption{Variation of $(1/N)dN_{d}/dp_T$ for $c{\bar{c}}$ (solid) and $b{\bar{b}}$ (dashed) versus their dissociation momenta at fixed evolution time, $\tau=1.0$ fm/c.}
		\label{Fig:spectra_10}
	\end{figure}

Following our examination of the influence of Glasma evolution time and cutoff distance on the created pairs, our focus shifts to understanding the momentum evolution of the pairs just before their dissociation. We initialize the momentum of charm and beauty quarks using the prompt momentum distribution provided in Eq. \eqref{eq:FFNLO}. The momentum of their anti-particles is obtained through the relation ${\bf p}_{\bar q}={\bf p} - {\bf p}_q $, and with this, we let the pairs evolve within the strong and highly non-equilibrium color fields of Glasma. We consider two extremes of Glasma evolution time: $\tau = 0.5 $ fm/c and  $\tau = 1.0 $ fm/c, presented in Figs. \ref{Fig:spectra_05} and \ref{Fig:spectra_10}, respectively. We observe the momentum of the pairs at the point of dissociation by noting when their separation reaches the specified cutoff distance, $r_c$, and plot the results against $\frac{1}{N}\frac{dN_d}{dp_T}$ ($N$ being the total number of initially created pairs ). The quantity $\frac{1}{N}\frac{dN_d}{dp_T}$ serves as a measure of the dissociated pairs.  Given our initialization, it is evident that most particles are produced within the range $p_T \in (0, 5)$ GeV. Consequently, the dissociation peaks in both figures are observed within this range.

In Fig.\ref{Fig:spectra_05}, we select different separation cutoffs, $r_c =$ 0.5 fm (red curve), 0.6 fm (blue curve), 0.7 fm (green curve), for an evolution time of $\tau=0.5$ fm/c. Solid curves represent $c {\bar c}$ results, while dot-dashed curves represent $b {\bar b}$. Two key points emerge from this analysis: (i) the peak of pairs dissociation and (ii) the shift in pairs momenta, $p_T$. The tallest peak is observed at the small $r_c = 0.5$ fm, which subsequently diminishes as we increase the cutoff distance. This aligns with our earlier discussion—the distance between pairs increases from their initial separation, and hence, pairs achieve the shortest distance first. Furthermore, as expected, the peak of $c {\bar c}$ is taller than that of $b {\bar b}$. Moreover, an increase in the cutoff distance results in a subtle shift of the peak towards higher $p_T$. This suggests that beyond individual quarks gaining a small yet discernible amount of energy \cite{Ruggieri:2018rzi}, the pair as a whole also experiences an energy gain within the Glasma. In simple words, the pairs reaching higher cutoff distances (which means individually going in different directions from each other despite the presence of {attractive potential}) require more energy to shift toward higher momentum. However, the number of pairs gaining more energy decreases, as evident from the reduced peak. 
%In the case of $b{\bar b}$, the forward shift in spectra is slightly less pronounced than in the $b{\bar b}$ case, indicating fewer interactions. 
To validate our observations, we calculate the area under each curve, obtaining precise numbers as shown in Fig.\ref{Fig:diss_vs_t}. 

Similar analyses are presented in Fig.\ref{Fig:spectra_10} for a higher evolution time, $\tau=1.0$ fm/c. We observe an increase in peak for all cases, intuitively understood as, given more time, more pairs reaching the cutoff distance with a decreasing trend, consistent with Fig. \ref{Fig:spectra_05}. Here, the difference in $b {\bar b}$ and $c {\bar c}$, peaks are more visible at a particular cutoff distance and all the peaks are shifted to slightly higher momentum due to the fact that they gain more energy in a longer time.  A noteworthy observation is that the $b {\bar b}$ spectra shifted more towards the higher momenta compared to $c {\bar c}$. One potential reason could be that the beauty quarks experience higher momentum broadening in the dense gluon fields, hence, they diffuse more, resulting in their broader spectrum.

\begin{figure}[t!] 
		\includegraphics[scale=.5]{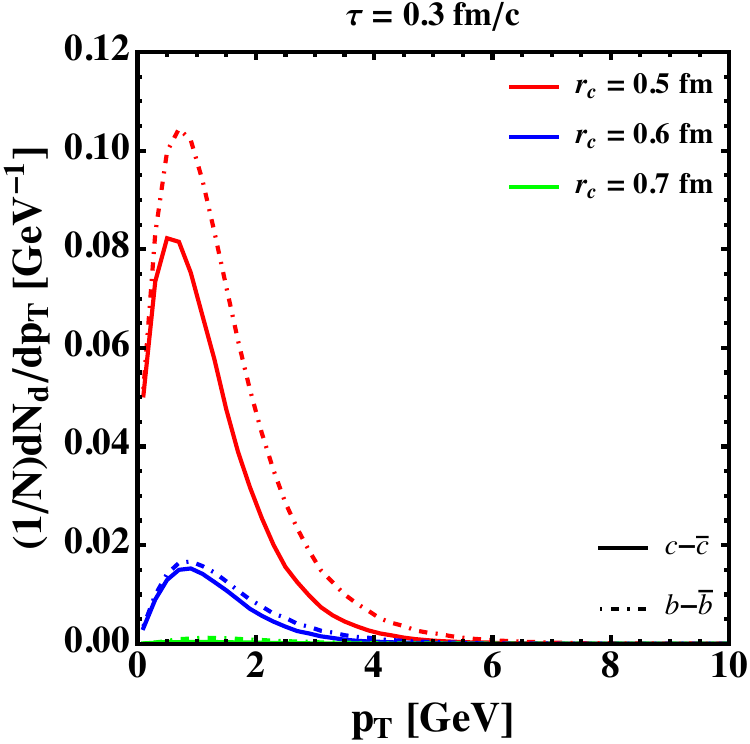}
		\caption{Variation of $(1/N)dN_{d}/dp_T$ for $c{\bar{c}}$ (solid) and $b{\bar{b}}$ (dashed) versus their dissociation momenta at fixed evolution time, $\tau=0.3$ fm/c.}
		\label{Fig:spectra_03}
	\end{figure}

In	Fig.\ref{Fig:spectra_03}, we again plot momentum dissociation spectra for $c \bar c$ and $b\bar b$ for $r_c =$ 0.5 fm (red curve), 0.6 fm (blue curve), 0.7 fm (green curve), for an evolution time of $\tau=0.3$ fm/c. Here, the peaks exhibit significant suppression, clustering around $p_T \approx 1.0$ GeV. There is no noticeable shift of the peak towards higher $p_T$ with an increase in the cutoff distance. Results at $r_c = 0.7$ fm are nearly completely suppressed, whereas at $r_c = 0.5$ fm, they reach a peak value of approximately $\sim 0.1$ GeV$^{-1}$, which is roughly one-third of both $\tau=0.5$ fm/c and $\tau=1.0$ fm/c suggesting a very small dissociation of $c \bar c$ and $b\bar b$ pairs. Conversely, at such a shorter time scale of $\tau=0.3$ fm/c, the results for $b\bar{b}$ dominance over $c\bar{c}$ are evident, consistent with the findings depicted in Fig. \ref{Fig:diss_vs_r}.

{\color{black}
\subsection{Estimate of the effect of the octet repulsive potential}
So far, we focused on color singlet quark-antiquark pairs. 
Differently from recent studies that used the Lindblad equation for studying 
the same problem in QGP \cite{Brambilla:2016wgg, Miura:2022arv, Casalderrey-Solana:2012yfo}, 
we have not implemented the transition to the octet
states.
In fact, these transitions are possible
in the pre-equilibrium stage as well, 
as the result
of the interaction of the quarks
in the pair with the Glasma fields.
Since we do not have wave functions describing the quark-antiquark pairs, rather we have 
an ensemble of classical color charges, the projection into subspaces of octet and singlet states
should be properly defined
within our approach. 
A possibility would be to decompose the tensor
$Q_a \bar{Q}_b$, where $Q_a$ denotes the color charge
of the quark and $\bar{Q}_b$ that of the antiquark,
into its singlet and octet components.
A detailed analysis of this important point is beyond the scope of our work
and will be presented elsewhere. However,
we can estimate the time scale at which this transition
should be important.

Firstly, at the initial time,
the pairs are in the color-singlet state. 
Specifically, let \( Q_a \) and \( \bar{Q}_a \) denote the color charges of one quark and its companion antiquark, where \( a \) represents the standard adjoint color index, \( a = 1, \dots, N_c^2 - 1 \). We initialize the charges with the condition that the net color charge in each pair and each \( a \) vanishes, \( Q_a(\tau = 0) + \bar{Q}_a(\tau = 0) = 0 \). Thus, at \( \tau = 0 \), the net color charge carried by the pair is zero. Additionally, at the initial time, \( Q_a \) and \( \bar{Q}_a \) are maximally anti-correlated. We then introduce the gauge-invariant color charge correlator,
\ba
G(\tau) \equiv -\sum_a \langle Q_a(\tau) \bar{Q}_a(\tau) \rangle,
\ea
where the average is taken over the ensemble of gluon fields (i.e., averaging over many events in the simulation). The overall minus sign in \( G(\tau) \) is added for convenience. It is reasonable to assume that the chance for the pairs to fluctuate into the octet state increases if \( G(\tau) \) decreases from its original value during evolution.

\begin{figure}[h!]
\begin{center}
\includegraphics[height=6.5cm, width=8.2cm]{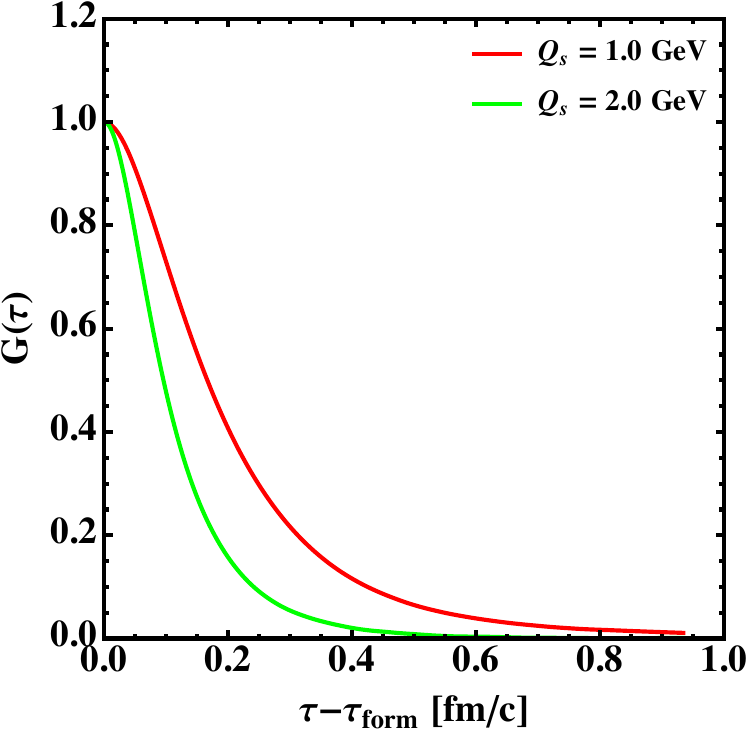}
\caption{ \( G(\tau) \) versus \( \tau \) for two representative values of \( Q_s \). }\label{fig:aaattt}
\end{center}
\end{figure}

Hence, we computed \( G(\tau) \) versus \( \tau \). The result of this calculation is shown in Fig.~\ref{fig:aaattt} for two representative values of the saturation scale, \( Q_s \). In the figure, \( \tau_\mathrm{form} \) corresponds to the formation time of the heavy quarks. We notice that \( G(\tau) \) decreases with \( \tau \). As evolution proceeds, correlations among the color charges of the quark and antiquark in the pair degrade, and it is reasonable to assume that the pairs will have the chance to fluctuate into an octet state. Based on the results shown in Fig.~\ref{fig:aaattt}, we estimate the decay time of the correlator, \( \tau_\mathrm{decay} \), by choosing, for example, the value of \( \tau - \tau_\mathrm{form} \) such that \( G(\tau_\mathrm{decay}) = 1/e \approx 0.368 \). We find that \( \tau_\mathrm{decay} \approx 0.2 - 0.3 \) fm/c. Thus, \( \tau_\mathrm{decay} \) is very close to the typical lifetime of the pre-equilibrium, gluon-dominated stage, which is the only stage where our model is phenomenologically relevant (we also show results for higher values of the proper time for illustrative purposes). Therefore, based on this analysis, we expect that the number of pairs fluctuating to the octet channel is not large.

Nevertheless, we prepared a set of simulations in which we switched off the color-singlet potential and introduced a color-octet, repulsive potential. This simplified model allows us to estimate the range in which the results would fall if we properly included both color-singlet and color-octet states. In these simulations, we used a perturbative form of the octet potential \cite{Singh:2018wdt},
\ba
V(r) = \frac{3}{4} \frac{\alpha_s}{3r},
\ea
where the overall \( \frac{3}{4} \) is the Casimir invariant \( C_F = (N_c^2 - 1)/2N_c \) of \( SU(2) \) (our simulations use two colors for simplicity), and we used \( \alpha_s = 0.433 \) for \( Q_s = 1 \) GeV and \( \alpha_s = 0.303 \) for \( Q_s = 2 \) GeV, as used for the singlet case.

\begin{figure}[h!]
\begin{center}
\includegraphics[height=6.5cm, width=8.2cm]{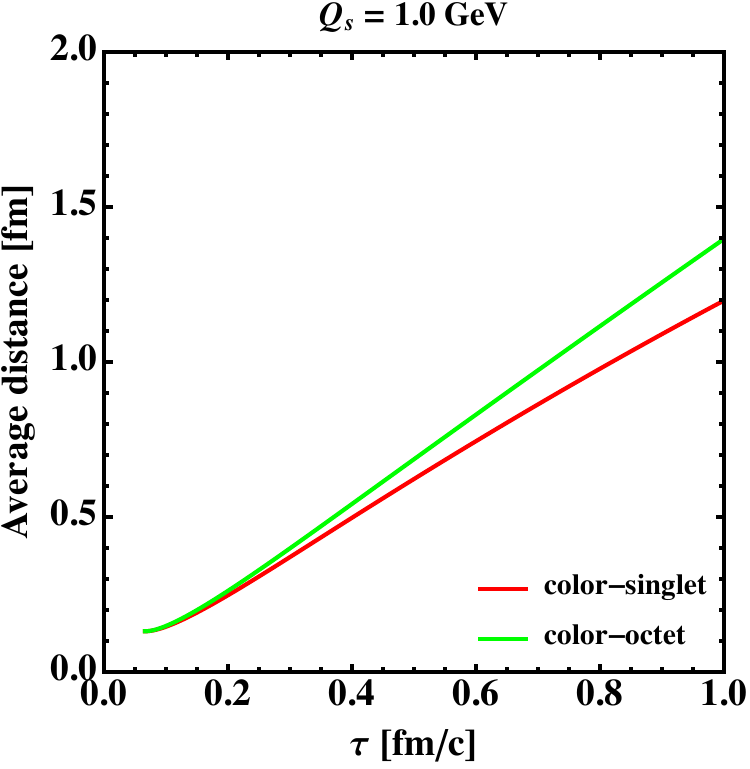}\\
\includegraphics[height=6.5cm, width=8.2cm]{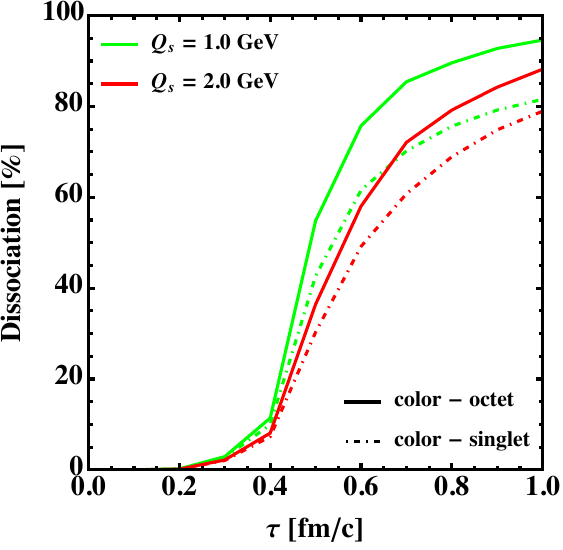}
\caption{
Upper panel: separation between \( c \) and \( \bar{c} \) for \( Q_s = 1 \) GeV. Lower panel: dissociation percentage versus time for \( c\bar{c} \) pairs.}
\label{TripletPotential_charm_r_vs_t}
\end{center}
\end{figure}

The results of these calculations are shown in Fig.~\ref{TripletPotential_charm_r_vs_t}. 
In the upper panel of the figure,
we show the evolution of the average separation between \( c \) and \( \bar{c} \) for the color-singlet and color-octet potentials. As expected, the effect of the octet potential is to accelerate the spreading of the pairs in coordinate space due to the repulsive interaction between the quark and the antiquark. While the color-singlet potential tends to keep the pair compact in contrast to the melting induced by the background fields, the color-octet potential facilitates the destruction of the pairs. In the lower panel of Fig.~\ref{TripletPotential_charm_r_vs_t}, we plot the dissociation percentage of \( c\bar{c} \) pairs versus time for two values of \( Q_s \), for both color-singlet and color-octet potentials. We find that, for a given \( Q_s \), the dissociation percentage increases when using the color-octet potential. This agrees with our previous discussion and is due to the repulsive nature of the potential. However, up to \( \tau \approx 0.5 \) fm/c, the discrepancy between results obtained with the singlet and octet potentials is less than 10\%.  This analysis allows us to state that
within the lifetime of the pre-equilibrium, gluon-dominated stage of the collision, fluctuations from color-singlet to color-octet states, although possible, might not substantially alter our predictions for the dissociation rate.
We acknowledge, however, that a more quantitative study should be done in this direction,
and this will be the subject of future works.
}

\section{Summary and Conclusions}
\label{sec:Summary_Conclusions}
In conclusion, our research delves into the evolution and dissociation dynamics of $q \bar{q}$ pairs within the Glasma medium slightly bound by the {attractive potential} a bit modified to regulate divergence, ensuring numerical stability. We observe that despite the application of an {attractive potential}, the separation between $q \bar{q}$ pairs increases due to the dominant influence of strong Glasma fields. This leads to the dissociation of pairs, with the dissociation percentage ranging from 0\% for $\tau = 0.5$ fm/c and $r_c = 1$ fm to approximately 80\% for $\tau = 1.0$ fm/c and $r_c = 0.6$ fm. Next, drawing attention to nuanced differences between $c\bar{c}$ and $b\bar{b}$ pairs, our results illuminate the role of quark flavor in the dissociation process within the Glasma. Notably, $b\bar{b}$ pairs exhibit a dissociation pattern similar to $c\bar{c}$ but with a slightly lower magnitude.

This distinction prompts further exploration into the specific interactions or dynamics experienced by heavier quark flavors within the Glasma. Examining dissociation momenta with the corresponding $\frac{1}{N}\frac{dN_{d}}{dp_T}$ provides insights into the energy/momentum distribution of dissociated pairs.
Analyzing the dissociation spectra, we observe that the dissociation peak is influenced by the chosen cutoff distance, with the tallest peak occurring at the lowest chosen cutoff, $r_c = 0.5$ fm. The shift in the peak towards higher momentum indicates that not only individual quarks gain energy, but the pair as a whole gains energy in the Glasma. This shift is more pronounced in charm than in beauty due to the latter's heavier mass and fewer interactions with the Glasma. Studying dissociation spectra at different Glasma evolution times reveals an increasing peak, emphasizing the impact of evolution duration on the dissociation process. At $r_c \ge 0.9$ fm and $\tau = 0.6$ fm/c, both $c \bar{c}$ and $b \bar{b}$ results overlap, suggesting a negligible mass effect at short evolution times. The observed dissociation percentage variations at different values of  $\tau$ and $r_c$ highlight the sensitivity of the process to these parameters. {\color{black}Later, we also presented a detailed discussion on the possible effect of considering the octet repulsive potential in the current analysis.}

This study contributes insights into the complex dynamics of $q \bar{q}$ pairs in the Glasma medium, shedding light on their dissociation mechanisms and momentum evolution. However, being a first step in this direction, it opens many possibilities for improvement. Refinements in the initialization process and a better understanding of $q \bar{q}$ dissociation could better estimate the amount of quarkonia, D, and B mesons entering the QGP medium produced in HICs. One of the refinements in the initialization of the position of quark and antiquark could be through the consideration of the wave function of the positronium system. However, this is not within the scope of the current analysis, and we leave it for future investigations. Moreover, exploring this setup of dissociation in the expanding bulk would make it more realistic. Furthermore, combining this study with QGP analysis, incorporating recombination possibilities ~\cite{Singh:2023zxu}, may enhance the understanding of experimental results.

{\color{black} The existence of screening in the Glasma is a topic of debate \cite{Boguslavski:2021kdd}. However, screening effects may occur near the equilibrated plasma or QGP. Since our analysis is specifically focused on the Glasma state and does not extend to the QGP medium, we have not incorporated these effects into our study. Our primary objective is to assess whether the quark-antiquark pair dissociates within the Glasma state. The consideration of screening effects becomes particularly important when transitioning from the Glasma to the QGP. This study lays the foundation for understanding dissociation dynamics in the Glasma state and can be further refined to include screening effects as the system evolves into the QGP phase.}	
	
\begin{acknowledgements}
SKD and PPB acknowledge the support from DAE-BRNS, India, Project No. 57/14/02/2021-BRNS. MYJ would like to acknowledge the SERB-NPDF (National postdoctoral fellow) fellowship with File No. PDF/2022/001551. This work has been partly funded by the
European Union – Next Generation EU through the
research grant number P2022Z4P4B “SOPHYA - Sustainable Optimised PHYsics Algorithms: fundamental
physics to build an advanced society” under the program
PRIN 2022 PNRR of the Italian Ministero dell’Università
e Ricerca (MUR).
\end{acknowledgements}

	% xxxxxxxxxxxxxxxxxxxxxxxxxxxxxxxxxxxxxxxxxxxxxxxxxxxxxxxxxxxxxxxxxxxxxxxxxx    

\end{document}